# Challenges for regional innovation policies in CEE countries: Spatial concentration and foreign control of US patenting

Balázs Lengyel[1,2*], Tamás Sebestyén[3], and Loet Leydesdorff[4]

[1] International Business School Budapest, Tárogató út 2-4, 1021 Budapest, Hungary. Email: blengyel@ibs-b.hu; phone: +3630-437-7807; fax: +36 1 391 2550

[2] Centre for Economic and Regional Studies, Hungarian Academy of Sciences, Budaörsi út 45, 1112 Budapest, Hungary. Email: lengyel.balazs@krtk.mta.hu

[3] MTA-PTE Innovation and Economic Growth Research Group, University of Pécs, Rákóczi út 80, 7622 Pécs, Hungary

[4] Amsterdam School of Communication Studies (ASCoR), University of Amsterdam, Kloveniersburgwal 48, 1012 CX Amsterdam, The Netherlands. Email: loet@leydesdorff.net.

Total word count: 7,145

**Funding**

This work was supported by the Hungarian Scientific Research Board [PD-106290 to B.L.] and IBS Research Grant.

**Acknowledgements**

We thank Klaus Körmendi and two anonymous referees for comments on a previous draft.

[*] Corresponding author.

**Challenges for regional innovation policies in CEE countries:**
**Spatial concentration and foreign control of USPTO patents**

## Abstract

Using techniques of data collection and mapping as overlays to Google Maps—on the basis of patent information available online at the U.S. Patent and Trademark Office (USPTO)—we point at two major and interconnected challenges that policy-makers face in Central and Eastern Europe (CEE) when combating the lagging innovation performance. First, we address the spatial concentration by using a distribution analysis at the city level. The results suggest that patenting is concentrated in post-socialist territories more than in western nations and regions. However, there is not a single outstanding hub in CEE when one compares USPTO patents normalized for the respective population sizes. Secondly, we argue that dominance of foreign control over USPTO patents is mostly embodied in international co-operations at the individual level, and only rarely spilled-over to MNE subsidiaries. In our opinion, catching-up of CEE in terms of patenting is unlikely, unless innovation policy measures focus on growing hubs and target both domestic inventors and international relations of companies.

## Introduction

Although economic growth of Central and Eastern European countries (CEE) has not been led by innovation in recent times (Varblane *et al.* 2007), much has been done regarding innovation policies in these countries during the past two decades (Kattel *et al.* 2009). Despite all the efforts, the complex issues of economic transition and simultaneous appearance of MNEs could not have been sufficiently countered by CEE national innovation policies (Inzelt 2008; Lengyel and Cadil 2009; Radosevic 2011). Furthermore, the heritage from the previous regime in the science and technology systems and foreign takeovers affected the regional reorganization of innovation systems at the same time (Blazek and Uhlir 2007; Lengyel and Leydesdorff 2011; Radosevic 2002).

Several scholars warned that the EU innovation policies were appropriate for conditions of core countries in the EU, but will probably not work in CEE (von Tunzelmann and Nassehi 2004). One can hardly find any specific innovation measures focusing on accession countries in the current EU policy documents. Even if policies aim at making regions that lag behind close up, stated objectives such as achieving globally competitive output in general will remain ineffective without further specification (EC 2011). In our opinion, a more detailed analysis is needed into CEE trends in order to ease the sharp divides of innovation outputs between EU core and accession countries (EC 2009).

Our argument is empirically based on data collected from the US Patent and Trade Office (USPTO) with at least one inventor address in the following Central and Eastern European countries: Czech Republic, Hungary, Poland, and Slovakia. USPTO data have been chosen instead of EPO data for two reasons. First, the accession of CEE countries to the common EU market makes the analysis of EPO patenting over the post-socialist period difficult, because the accession of CEE countries to the EU and common market have affected the number of EPO patent applications for reasons other than inventions (Hall and Helmers 2012). Second, USPTO patents can be expected to capture globally competitive innovation output better than EPO data, because USPTO patents provide more efficient protection than EPO patents (Ginarte and Park 1997, Martinez and Guellec, 2003).

We visualize spatial trends of patenting during the period 1980-2010 and analyse this data more deeply for the year 2007 and thereafter. For the purpose of a further analysis, we retrieved also German patents published in 2007 in order to broaden the comparison to the former GDR and its western counterpart. The two central arguments will be about the geographical concentration of patenting and the conditions of foreign control and local spill-over.

First, figures drawn from patent statistics at the regional level may fail in informing regional innovation policies because administrative regions are larger in CEE countries than in core countries of the EU. Thus, a city-level investigation is necessary to monitor spatial trends in CEE patenting performance. Indeed, when analysing USPTO maps at the city level, one finds a geographically concentrated pattern of patenting more in areas that have gone through economic transition than in regions and countries where the market economy has developed more gradually in terms of its knowledge base. Unlike in previous East-German states (Franz 2010; Fritsch and Graf 2011), however, there are no real hubs in CEE that could power a catching-up processes. Focusing on hub creation might be a way to boost regional innovation output.

Second, foreign control over USPTO patenting in CEE countries might hinder those innovation policy measures that do not take the type of foreign control into consideration. Our findings suggest that the vast majority of CEE inventors take part in international projects centred at foreign locations. This pattern is very different from patent collaborations in EU15 countries that are regionally or nationally bounded (Cheesa *et al.* 2013). One can assume that the local effect of the innovation process is weaker with an assignee abroad than with a local assignee (be it an MNE or a domestic firm). Therefore, one cannot expect patent concentrations to result in a catching-up process similar to the ones in East-Germany unless the different types of foreign control are addressed by deliberate policies.

The remainder of the paper is structured as follows. Innovation trends and innovation policy failures during the post-socialist transition are first summarized in the next section. The third section introduces the data collection and mapping techniques. The city-level distribution of the USPTO patents is analyzed in the fourth section; in section five we discuss the types of foreign control over CEE patenting. In the final section, further research questions and policy implications are elaborated.

**Post-socialist transition, innovation policy failures, and lags in patenting**

The CEE countries entered a transition period from a centrally planned economy to a market economy after the fall of the Berlin Wall in 1989 and the demise of the Soviet Union in 1991. The transition and thereafter the accession to the EU, however, faced the challenges of globalisation during the same period of time (Enyedi 1995). Consequently, the ‘crescendo’ of global research networks and the transition in Central-European innovation systems operated in parallel. Multinational enterprises (MNEs) became the major actors in these countries (Radosevic 2002; UNCTAD 2005): business R&D was integrated into global networks and local portfolios were streamlined in terms of possible overlaps. Economic transition has left a footprint on regional dynamics as well, since foreign direct investments (FDI) favoured metropolitan and Western locations (Petrakos 2001). However, universities and state-controlled services could maintain an organizing role in the innovation systems of regions lagging behind (Lengyel and Leydesdorff 2011).

The formation of institutional arrangements did not take place at the same pace, but lagged (Freeman and Perez 1988). Innovation policy suffered from major fallbacks in R&D that followed on EU harmonization. Some authors nevertheless signalled the awakening of national innovation systems (e.g., Suurna and Kattel 2010). However, the national innovation policies of CEE failed in meeting the above mentioned challenges of transition, accession, and globalization (Havas 2002; Radosevic and Reid 2006; Tiits et al. 2008). For example, university-industry relations were not established at a sufficient level (Inzelt 2004). Furthermore, although MNEs can be expected to have distinctive power in shaping national S&T policies (Inzelt 2008), CEE countries were late in targeting large foreign-owned companies (Lengyel and Cadil 2009) and failed in encouraging the link creation between these foreign and domestic companies (Radosevic 2011).

We use USPTO data in our analysis instead of EPO data, because this data is more reliable over the whole period of economic transition and because it captures globally competitive innovation (Patel and Pavitt 1995; Nagaoka *et al.* 2010). The latter feature of analysing innovation performance in CEE countries is crucial, because inventors in post-socialist countries tend to limit IP protection to their local markets. However, EPO and USPTO patenting are of the same order of magnitude. Data from the annual reports of the Hungarian, Polish, Czech, Slovak, and German national patent offices concerning the number of applications by domestic applicants,[1] are supplemented with patent applications to EPO and USPTO from these countries in Table 1.

Table 1

We added data for the Netherlands (as a country of comparable size) and Germany (as a world leading country in USPTO patents) to Table 1 in order to show the discrepancies. One can observe an order of magnitude higher number of patent applications to the national patent offices than to EPO or USPTO in the case of CEE countries. The same tendency is apparent in Germany, but the relative numbers are still less sharp when compared with data for the Netherlands with an internationally oriented economy. Note that minor differences may result from the different counting methods in various nations. Although EPO and USPTO applications vary from year to year between 2000 and 2010, national applications outnumber them during this whole decade.

Table 2

Table 2 summarizes the USPTO patenting activity during the 1980-2007 period for the selected CEE countries as well as Germany and the Netherlands. We calculated the average number of USPTO patents granted per year in five-year windows. These numbers reveal that patenting performance declined in the Czech Republic and Hungary in the last years of the socialist period and especially after 1990. The Czech system could catch-up to the level of patenting of the early ’80s only in the first half of the 2000s; whereas the numbers of USPTO patents granted per year had not yet reached the output of the ’80s in Hungary in 2007. Transition seems to have run its’ course faster in Poland and Slovakia. The latter country was established as an independent nation state after the socialist regime fell.) In comparison, the number of granted USPTO patents has increased gradually in Germany and the Netherlands.

[1] Applications to EPO broken down by the country of the inventors were drawn from Eurostat database. Data on USPTO patent applications is available at the USPTO webpage: http://appft.uspto.gov/netahtml/PTO/search-adv.html. An important difference between the USPTO and the EPO statistics is that the former counts applications with at least one inventor from the specific country whereas the latter provides a fractional count.

On an aggregate country level, accession countries have a more than two orders of magnitude lag in terms of patent applications submitted to the European Patent Office compared with EU15 countries (EC 2009). Despite of the sometimes high growth rates and some emerging regions, CEE countries struggle with even reaching the level of their patenting activity in the socialist era. We hypothesize two processes behind this phenomenon in this paper: regional restructurations and foreign takeovers. Let us, however, we introduce the data collection and visualization techniques that enable us to make the argument empirically.

## Mapping USPTO patents: data collection and visualization technique

We use techniques for patent retrieval and statistically informed mapping developed recently by Leydesdorff and Bornmann (2012), which can be repeated easily by anyone without any special knowledge in informatics or GIS. Let us introduce this method shortly before turning to our special case of CEE countries.

### *Data collection and mapping technique*

The database of the USPTO contains all patent data since 1790. Patents are retrievable as image files since then, and after 1976 also as full text. The HyperText Markup Language (HTML) format allows us to study patents in considerable detail (Leydesdorff 2004). A set of dedicated routines was developed, which can be downloaded by the user at http://www.leydesdorff.net/software/patentmaps/index.htm. This Web page also contains further instructions.

Before running these programs, the user first composes a specific search string at the 'Advanced Search' engine of the USPTO database of granted patents at http://patft.uspto.gov/netahtml/PTO/search-adv.htm or patent applications at http://appft.uspto.gov/netahtml/PTO/search-adv.html. Using these databases, one can, for example, search with names of countries, states, or city addresses in addition to the issue and/or application dates of the patents under study or classifications (Leydesdorff et al. in press). All patents contain city addresses of inventors and assignees. The routines sort these addresses into overlays to Google Maps that can be used in web-browsers. Furthermore, the data is organized into a relational database that can be used for statistical analysis.[2]

For this study, we use primarily the number of citations to each patent as a quality indicator, and the address information of inventors and assignees. The facility of GPS Visualizer at http://www.gpsvisualizer.com/ geocoder/ was used for the geo-coding of the addresses. After geo-coding the user is prompted with further questions that influence the eventual layout of the map.

Using colours similar to those of traffic lights, nodes of cities with patent portfolios statistically significantly below expectation in terms of citedness are colored (dark) red and cities with portfolios significantly above expectation (dark) green[3]. Lighter colors (lime green and red-orange) are used for cities with expected values smaller than five patents (which should not statistically be tested); light green and orange are used for non-significant scores above or below expectation. The precise values are provided in the descriptors, which can be made visible by clicking on the respective nodes in the online version of the maps; see for example at http://www.leydesdorff.net/cee/top25.htm. Additionally, all numerical values are stored in the database file "geo.dbf" for statistical analysis.

A second output file ('patents.txt'; at http://www.leydesdorff.net/cee/patents.htm) contains the information for generating a map that is not based on citations, but on the portfolio of the patents harvested with the search. Cities are here compared in terms of numbers of patents. This representation focuses on geographical effects, such as agglomeration and diffusion, more than on the citation dynamics. To that end, a quantile—that is, the continuous equivalent of a percentile—is computed for each city by dividing the number of cities with fewer patents (than the city under study) as the numerator by the total number of cities in the set as the denominator. Using the same colours as Bornmann, Leydesdorff, Walch-Solimena, and Ettl (2011), the top 1% cities are colored red (as 'hot spots'), the top 5% fuchsia, the top 10% pink, the top 25% orange, the top 50% cyan, and the remainder (bottom 50%) is colored blue. These percentile rank classes follow the categorization used in the *Science and Engineering Indicators* series of the National Science Board (2012; cf. Bornmann & Mutz, 2011).

---

[2] Note that inventor and location names appear as transcriptions into English in the USPTO documents and therefore special (e.g., diacritical) characters from other languages can be misformed.

[3] For each city in the downloaded set, the observed number of these highly cited patents is statistically tested against the expectation using the *z*-test for two independent proportions (Sheskin 2011, p. 656). Statistical tests will only be performed for cities with expected values higher than 5 because the *z*-test (like the chi-square) is not reliable for expected values lower than five. Significance levels are indicated (in the clickable descriptors of the cities on the map) as follows: * for $p < 0.05$, ** for $p < 0.01$, and *** for $p < 0.001$.

*Hungary compared*

Our research was triggered by the surprising finding that the search string 'icn/nl and isd/2007$$'—in other words: inventor country Netherlands, and issue date in 2007—recalled 1,908 patents (Leydesdorff and Bornmann 2012), whereas the same search string for Hungary ('icn/hu and isd/2007$$') resulted in only 72 patents with inventors located in Hungary. We used 2007 in order to have a sufficiently long citation window for the discrimination. The difference by more than an order of magnitude resounded with our background knowledge about these two, otherwise comparable countries, and thus we decided to investigate this more intensively.

The spatial distribution of patents in Hungary is skewed: only 13 cities are listed among the addresses of two or more patents filed in 2007, whereas 128 cities were found in the Netherlands with five or more patents.

Figure 1

Inventors residing in Budapest, the capital of the country and the location of most of the multinational research labs, authored or co-authored 37 patents (Figure 1). The capital is followed by two university towns in terms of number of patents: Szeged in the South-East with seven patents and Debrecen in the East with six patents. Further towns with two or more patents are Miskolc (in the North-East), a centre for heavy industry and mining, and Székesfehérvár (South-West from Budapest) as a previous location of IBM. Some inventors were also housed in small agglomeration towns near Budapest; however, these inventors probably commute to the city. Interestingly, the North-Western part of the country that is considered as the most developed part in terms of the distribution of capital and where the automotive industry resides, does not show up as an important patenting area on the map.

Budapest, Miskolc, Szeged, and two small towns in the Budapest agglomeration are above the expectation in terms of number of patents filed (Figure 1a). Budapest was in the top 10%; Debrecen, Szeged, and Szombathely among the top 25%; Vác (North from Budapest) in the top 50%; all the other locations are in the bottom 50% in terms of patent portfolio. None of these results are statistically significant due to the low numbers of patents.

**Geographical concentration without hubs**

In order to visualize the strong geographical concentration and the lack of innovation hubs in CEE, data have been collected following the previous methodology. We add an asymmetry analysis of the patenting distribution to the empirical sections of the paper in which we compare maps at the city-level using USPTO data for Germany and the selected CEE countries in the year 2007 with the map of EPO data at the level of NUTS-3. [4] Thereafter, the spatial distribution of USPTO patents in West- and East-German states is compared with CEE data. Finally, maps with five-year time windows on CEE data can show the spatial evolution of USPTO patenting during the post-socialist transition.

*Patent maps in Germany and CEE*

We collected USPTO patents with inventors in Germany, the Czech Republic, Poland, Slovakia, and Hungary using the search string 'icn/(de OR cz OR pl OR sk OR hu) and isd/2007$$' on July 18, 2012. This recalled 11,178 patents. All inventor addresses were used to create overlays. In order to keep the maps readable, only cities with five or more patents are indicated (Figure 2). Note that Austrian cities are also marked in the map because of the addresses of co-inventors; but these locations are not used in the argument.

[4] NUTS is an abbreviation for "*Nomenclature des Unités Territoriales Statistiques*" (that is, Nomenclature of Territorial Units for Statistics). The NUTS classification is a hierarchical system for dividing up the economic territory of the EU.

Figure 2

Figure 2 shows that the vast majority of inventors are located in Germany, only very few cities had five or more patents in the other four countries. These locations include: Warsaw (21 patents) and Wroclaw (7 patents) in Poland, Prague (17 patents) and Brno (8 patents) in the Czech Republic, Bratislava (7 patents) in Slovakia, Budapest (37 patents), Debrecen (6 patents) and Szeged (7 patents) in Hungary. The *z*-test indicates patents from Budapest, Szeged, Warsaw, and Wroclaw to perform above expectations in terms of citations; but patents from Bratislava, Brno, Debrecen, and Prague are indicated as performing below the expected level. Due to the low number of patents in these accession countries, however, none of these results are significant ($p > 0.05$). The sharp difference in terms of patenting density between east and west is most pronouncedly visible in Figure 2. With the exception of the Berlin and Dresden areas, the figure shows the east/west divide still prevailing among the German States.

Figure 3

Regional patent statistics that contain EPO patent applications at the NUTS-3 regions level depict regional differences in Germany efficiently whereas this visualization might be misleading in CEE countries (Figure 3). For example, Figure 3 suggests that Czech regions perform at the same level regarding patent outputs and also indicates lagging regions in Hungary to perform at the same level as Budapest. We argue that city-level maps are more efficient in monitoring patenting performance in CEE for two reasons. First, NUTS 3 regions in CEE are larger in size than those in Germany (or in the Netherlands) and therefore might not allow for as precise representation as in EU core countries. Second, the differences among EU core and CEE patenting are too sharp to depict distributions efficiently; none of the CEE regions did exceed the upper limit of 50 patent applications per million inhabitants. Therefore, in order to understand what has happened to patenting over the post-socialist transition we need to analyse normalized distributions of city-level data.

*Spatial distribution of USPTO patents in CEE, 2007*

The data-collection and mapping technique presented above enables us to analyse patenting at the settlement-level. City-level comparison and spatial distribution analysis provides us with a more detailed tool than patent statistics (Deyle and Grupp 2005; Fornahl and Brenner 2009).

Previous research has found that the East-German closing-up process occurred in certain East-German cities ('cathedrals') only, whereas their surroundings have remained 'deserts' (Franz 2010; Fritsch and Graf 2011). In other words, the lower level of patenting in post-socialist areas is attached to a relatively stronger geographical concentration in specific cities than in regions that have developed more gradually.

Figure 4

The main difference in terms of spatial concentration of patenting activity across the countries covered by our data is in the relative performance of medium-sized locations. The rank-size distributions of USPTO patents from West-German, East-German States and CEE countries are illustrated in Figure 4.[5] Note that location size refers to the number of patents invented or co-invented there. One can construct power-law distribution in the West-German set of locations; however, a trend breach occurs after the first five ranks in the East-German set, and after the first three ranks in the CEE sets. Medium-size locations are represented to a lower extent in the two latter cases, which makes the distributions asymmetric. Consequently, our USPTO data analysis at the city level provides new evidence and confirms previous findings: patenting is more concentrated in post-socialist areas than in territories that have gone through a more gradual development.

A very interesting territorial correspondence can be revealed when controlling for the population size of cities in our data: the hubs that are expected to stand out from their environment do not stand out from the distribution, but both their West- and East-German counterparts do.

[5] Previously East-German States count for 2,952 cities or communes whereas there are 8,482 cities or communes in West-German States; this difference may influence patenting distribution. Data was downloaded from http://www.citypopulation.de/.

In order to illustrate the above statement we defined patenting intensity $PI_i^g$ as

$$PI_i^g = \frac{PAT_i^g / \sum_i PAT_i^g}{POP_i^g / \sum_i POP_i^g};$$

where $i$ denotes location in a group $g$. Groups are West Germany, East Germany, and CEE countries for reasons specified above. The idea behind is to compare the observed distributions to a theoretically even distribution of patents normalized for the respective populations. If $PI_i^g$ is larger than one, the given location contributes to patenting more than expected according to its population size; if $PI_i^g$ is smaller than one, patenting per inhabitant is less than the expected distribution suggests.

Figure 5 exhibits the relationship between the log of the population rank (in a descending order) and the log of the patent intensity (only locations that have at least one patent are included in the analysis). The horizontal line at the value of zero represents the even distribution; locations below the line have less patents per inhabitants than expected and locations above it outperform the expectation. Interestingly, smaller settlements seem to produce more patents per inhabitants than big settlements systematically, which –insofar as we know– has not been shown previously. However, we focus here on the discussion of the comparison among West-German, East-German, and CEE distributions.

Figure 5

The distribution of patenting intensity suggests that although the numbers of patents per inhabitants are smaller in large cities than in small towns, there are certain West German and East-German cities –called innovation hubs– that exceed the expected level of patenting intensity despite their relatively large populations (points above the line on the left side of the distributions). No such hubs were found in CEE countries. All the big cities in CEE perform below the expectation in terms of USPTO patenting except Budapest that fits exactly to the expectation.

These findings overshadow the catching-up chance of CEE regional innovation systems. Since the East-German closing-up was led by major innovation hubs such as Dresden and Jena, the lack of such hubs and the relatively weak performance of large cities indicates that the transformation process resulted in a different spatial distribution in CEE. The East-German development path is not yet followed.

Appendix 1 with six detailed maps over the 1981-2010 period supports these findings. Regional centres that stood out in the socialist era –like Brno in the Czech Republic, Katowice in Poland or Miskolc in Hungary– have lost their place after the system changed. Patenting became more concentrated in the capital cities during the ’90s. However, countries differ in terms of how regional centres have developed again into regional centres in the 2000s. In addition to the dominant agglomerations around capitals, Brno and Ostrava in the Czech Republic, Gdánsk and Bzeszów in Poland or Debrecen in Hungary seem to have an effect on their region. We suggest that policies for regional innovation could focus on these emerging hubs so that these centres can have a stronger impact on their environments.

**Foreign control and CEE patenting**

The takeover of innovation capacities by foreign companies during the post-socialist period in CEE countries has been defined as a major factor that innovation policies have to face (Blazek and Uhlir 2007; Inzelt 2008; Lengyel and Cadil 2009; Radosevic 2002; von Tunzelmann and Nassehi 2004; Varblane *et al.* 2007). Indeed, foreign ownership of USPTO patents that CEE inventors contributed to increased after 1990 (Figure 6). The pattern differs from that for countries like Germany or the Netherlands.

Figure 6

Policy-makers face a dilemma regarding foreign control on innovation activity. On the one hand, the gap between co-located foreign firms and domestic firms might be too sharp and therefore knowledge transfer is rare between them (von Tunzelmann and Nassehi 2004); MNEs are not easy to control by national policies and they have a strong hold on policy-making themselves (Inzelt 2008). On the other hand, foreign firms might bring new knowledge to a country and international collaboration –that is another form of foreign control– can be fruitful for knowledge creation, because the combination of distant knowledge bases may lead to more innovative outcomes (Maggioni et al. 2007; Sebestyén and Varga 2012).

We agree that one of the most important tasks of the innovation policies of CEE countries is to enhance the knowledge transfer between foreign and domestic companies (Radosevic 2011). However, one can argue that the type of collaboration determines the impact that innovation policies can have. For example, in the particular case of USPTO patenting, domestic CEE inventors who are controlled by foreign-owned assignees can collaborate with foreign partners (e.g. university professors participating in international projects) or might work for them (e.g. researchers in MNE subsidiaries). The first type of co-operation mainly remains at the individual level, while the second type of collaboration is controlled inside an organization (Guellec and van Pottelsberghe 2001).

USPTO patent data enable us to distinguish these two types of foreign control on the basis of inventors and assignee addresses. One can argue that individual collaboration in international inventor networks might be more difficult to enhance by policies than the organizationally controlled cooperation.

Table 3

In summary, 221 patents were counted in the four CEE countries in 2007, 50% of them were exclusively domestic. For a comparison: 83% of the German USPTO patents are exclusively domestic. There were more patents granted by the USPTO in 2007 that have been created in international collaboration than in cooperation between domestic inventors only in the Czech Republic, Poland, and Slovakia. However, Hungary seems to produce more patents locally than in international projects (Table 3). In summary, a large majority of Central European inventors co-authored international patents that have been submitted by foreign assignees. This CEE pattern is again very different from the German one.

A large share of the domestic assignees was local subsidiaries of MNEs in 2007; cross-country differences prevailed though. These companies outnumbered domestic firms in the Czech Republic and Hungary but domestically-owned firms were dominant in Poland and Slovakia (see Appendix 2).

Previous research has found that individual cooperation of university professors in international projects had hardly any local effect in Hungary (Lengyel *et al.* 2006). These individuals in the 'periphery' are intellectually motivated to enter international collaborations, but are often isolated from domestic partners (Goldfinch *et al.* 2003). Therefore, innovation policy should focus on transferring the knowledge gained in these international collaborations into a broader local community. Similarly, more attention is needed to support the learning process of domestic companies because one cannot expect automatic spillover from foreign companies (Békés *et al.* 2006; Lengyel and Leydesdorff 2013).

**Conclusions**

A fine-grained geographical mapping of USPTO patents was used in the above sections to illustrate the concentration patterns of innovation activities in Central and Eastern Europe and its lag from core countries in the EU. Concentration is stronger in CEE countries than in West Germany and seems to be similar to the spatial distribution in the former Eastern Germany. It could also be shown, however, that CEE centres have not yet evolved to such innovation hubs as centres in East Germany. Few regional centres, like Brno, Gdánsk or Debrecen seem to ease the dominating concentration around their capitals by slowly growing their surrounding environments into agglomerations. These results suggest that EU and national innovation policies to support these processes should focus on the creation of regional innovation hubs in order to put regional innovation of CEE countries on the path that East Germany is already walking.

Combating the huge lag in terms of innovation output cannot avoid a special focus on foreign-owned companies and international collaboration in R&D and patenting. The foreign control of USPTO patents is overwhelming and has been signalled already during the '90s (see

Figure 6) when the national systems of innovation suffered major breakdown in CEE. Foreign firms took over key sources of innovation capacities, but at the same time transferred new knowledge to these countries and the possibility for international collaborations opened up for CEE inventors. However, patenting remained at a very low level and there are hardly any domestic companies that apply for patents at USPTO. In our opinion, this indicates major obstacles for knowledge transfer between foreign and domestic stakeholders.

The local effects of individual inventors who participate in international projects tend to remain marginal from a structural perspective. Nevertheless, these collaborations are the primary source from which inventors have learned the ways how intellectual property is to be secured in the market economy and especially in very competitive markets. Similarly, the gap between foreign and domestic companies set innovation dynamics back. Automatic spill-over effects cannot be assumed from these collaborations to other local agents: innovation policies should focus on supporting these local effects of spill-over.

Very few articulations of innovation policies for CEE can be found in the current EU documents. For example, smart growth initiatives of the EU 2020 strategy do not deal with the internationalisation of innovation extensively. However, CEE countries seem to produce the majority of their globally competitive innovation in international collaborations (unlike Germany). Similarly, there are hardly any CEE examples among the cases of smart specialization. In our opinion, the creation of regional innovation hubs could be crucial for closing-up in Eastern Europe. To that end, EU innovation policy should pay more attention to the special needs of CEE countries and other new member states in these specific terms. National innovation measures can focus on creating regional innovation hubs and on strengthening the knowledge transfer from foreign-controlled innovation projects to the local communities.

Table 1: Number of patent applications by patent offices, 2007

| | National patent offices | EPO | USPTO |
|---|---|---|---|
| Czech Republic | 1706 | 182 | 196 |
| Hungary | 877 | 186 | 230 |
| Poland | 2996 | 200 | 135 |
| Slovakia | 240 | 37 | 67 |
| Germany | 47853 | 23907 | 22193 |
| Netherlands | 2078 | 3241 | 3902 |

Source: Self edited by the authors.

Table 2: Average number of granted USPTO patents per year in five-year periods, 1980-2010

| | 1981-1985 | 1986-1990 | 1991-1995 | 1996-2000 | 2001-2005 | 2006-2007 |
|---|---|---|---|---|---|---|
| Czech Republic | 48.69 | 24.496 | 22.682 | 38.122 | 61.758 | 60.93 |
| Hungary | 143.304 | 82.222 | 45.724 | 61.886 | 88.218 | 51.225 |
| Poland | 18.082 | 11.422 | 16.716 | 29.512 | 59.228 | 40.325 |
| Slovakia | 0 | 0 | 1.574 | 7.078 | 11.826 | 16.39 |
| Germany | 7909.638 | 7741.416 | 8205.018 | 12418.42 | 14118.02 | 8154.195 |
| Netherlands | 907.408 | 972.788 | 1032.372 | 1561.462 | 1930.69 | 1050.51 |

Source: Self edited from EUROSTAT data retrieved from
http://appsso.eurostat.ec.europa.eu/nui/show.do?dataset=pat_us_ntot&lang=en

Table 3: Foreign control of USPTO patents in four CEE accession countries and Germany, 2007

| | Czech Republic | Hungary | Poland | Slovakia | Germany |
|---|---|---|---|---|---|
| Number of patents | | | | | |
| in international co-operation | 32 | 29 | 43 | 8 | 1865 |
| exclusively domestic | 25 | 43 | 34 | 7 | 9103 |
| Number of inventors | | | | | |
| working for foreign assignees | 68 | 101 | 53 | 5 | 3001 |
| working for domestic assignees | 22 | 56 | 29 | 4 | 14792 |
| unknown | 12 | 9 | 14 | 4 | 425 |

Note: Inventors and assignees have been associated using patent IDs. Additional information on CEE assignees was collected by the authors.

**Fig. 1a:** Portfolio for cities in Hungary with two or more patents,

based on integer counting of the inventors

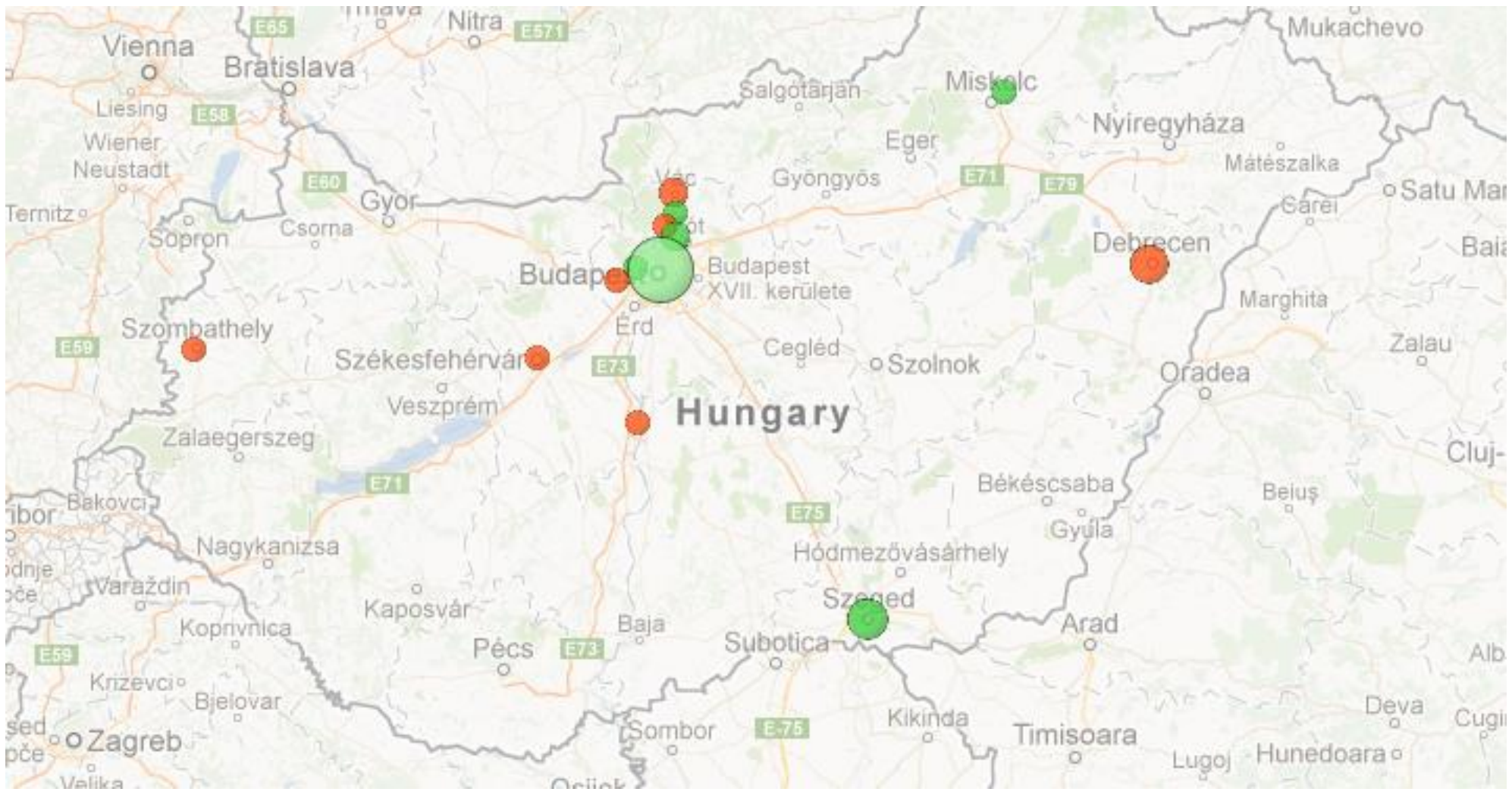

Note: The node sizes are proportionate to the logarithm of the number of patents. No scores are significant due to the low number of patents.

**Fig. 1b** Portfolio for cities in Hungary with two or more patents,

in terms of percentiles

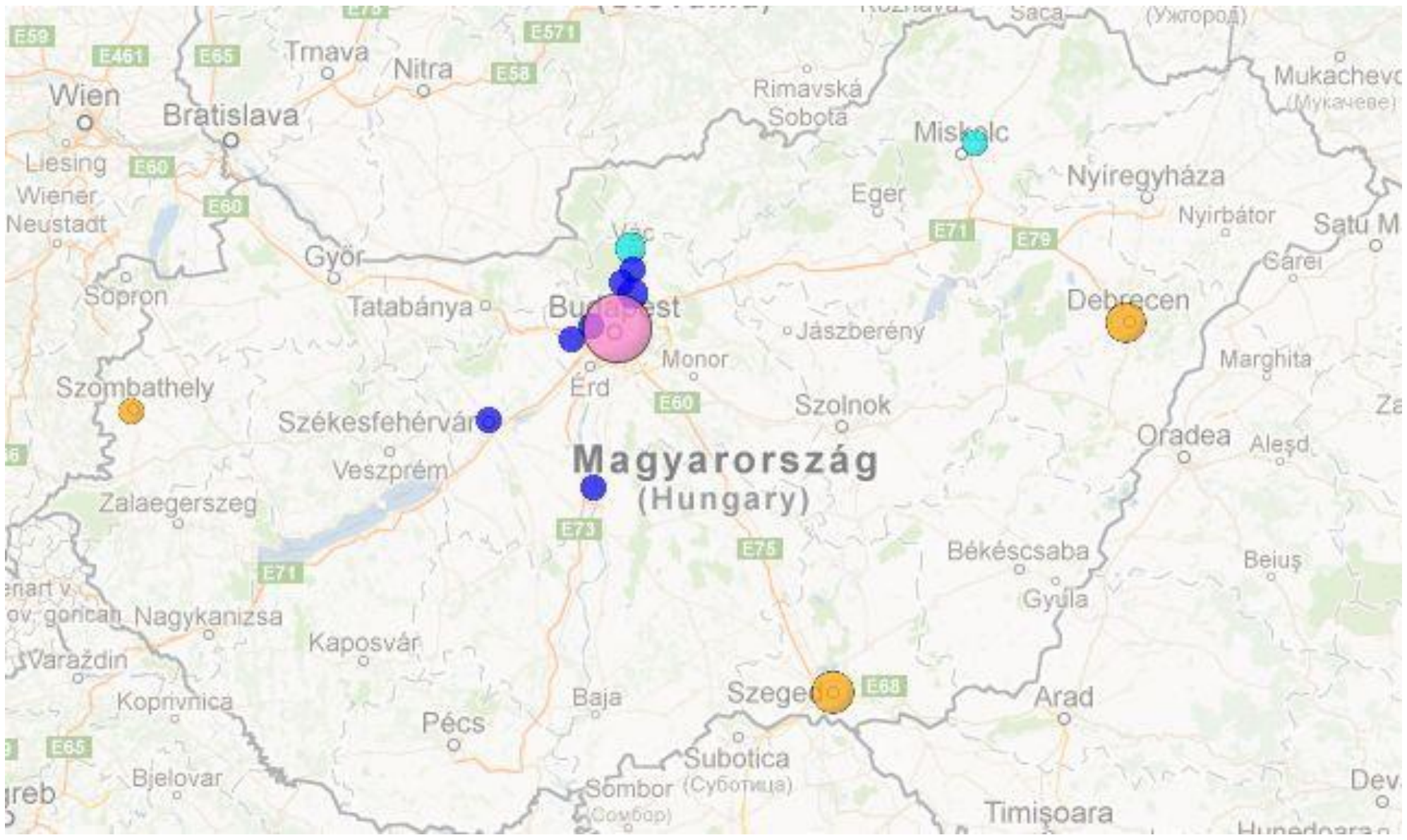

Note: Node colors represent percentiles: top 1% red; top 5% fuchsia; top 10% pink; top 25% orange; top 50% cyan; and bottom 50% blue. The node sizes are proportionate to the logarithm of the number of patents. No scores are significant due to the low number of patents.

**Fig. 2:** Portfolio for cities in Central and Eastern Europe with five or more patents, based on integer counting of the inventors

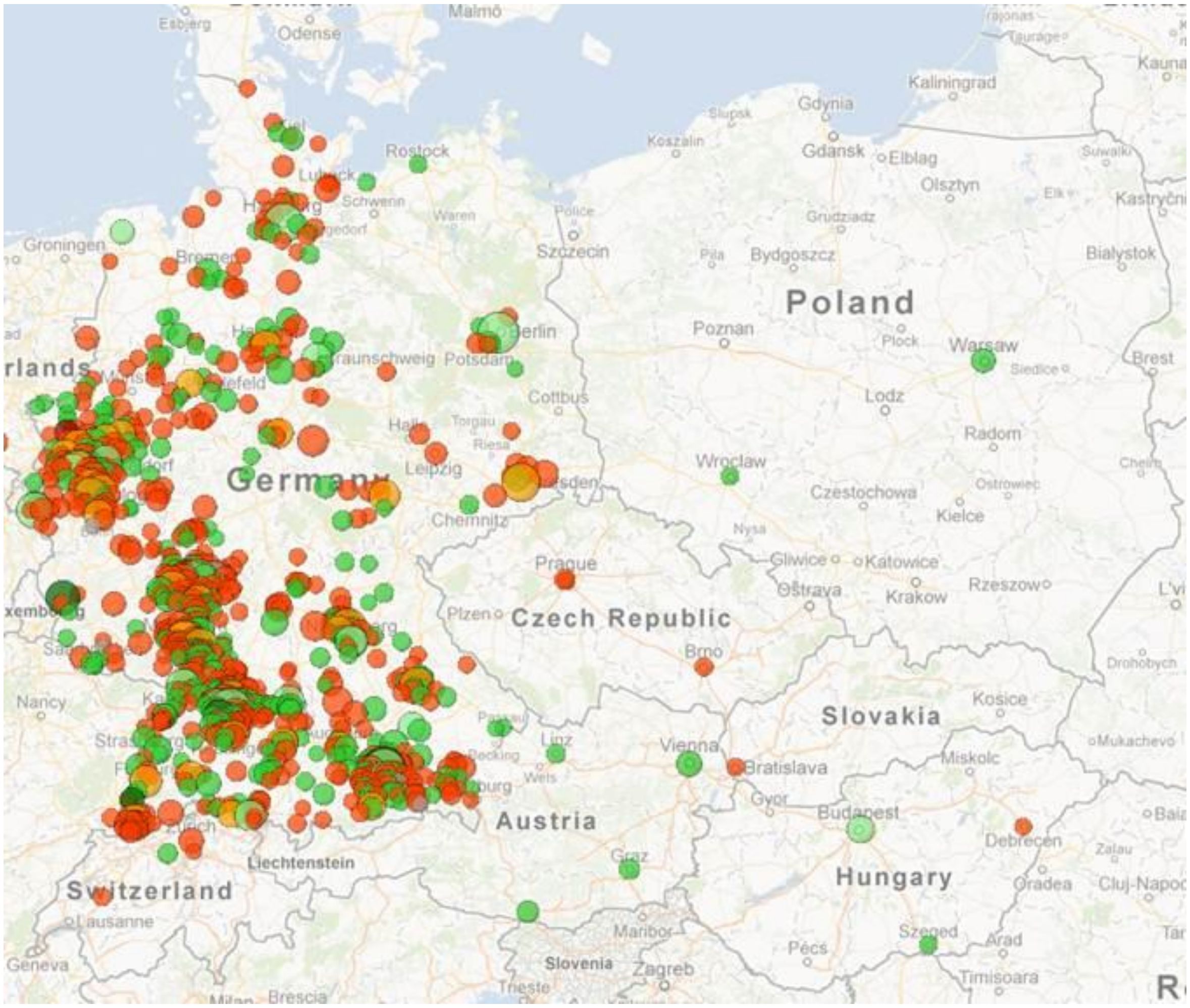

Note: The node sizes are proportionate to the logarithm of the number of patents. For an interactive version, see at http://www.leydesdorff.net/cee/top25.htm

**Fig. 3:** Patent applications from Central European NUTS 3 regions to the EPO, 2008

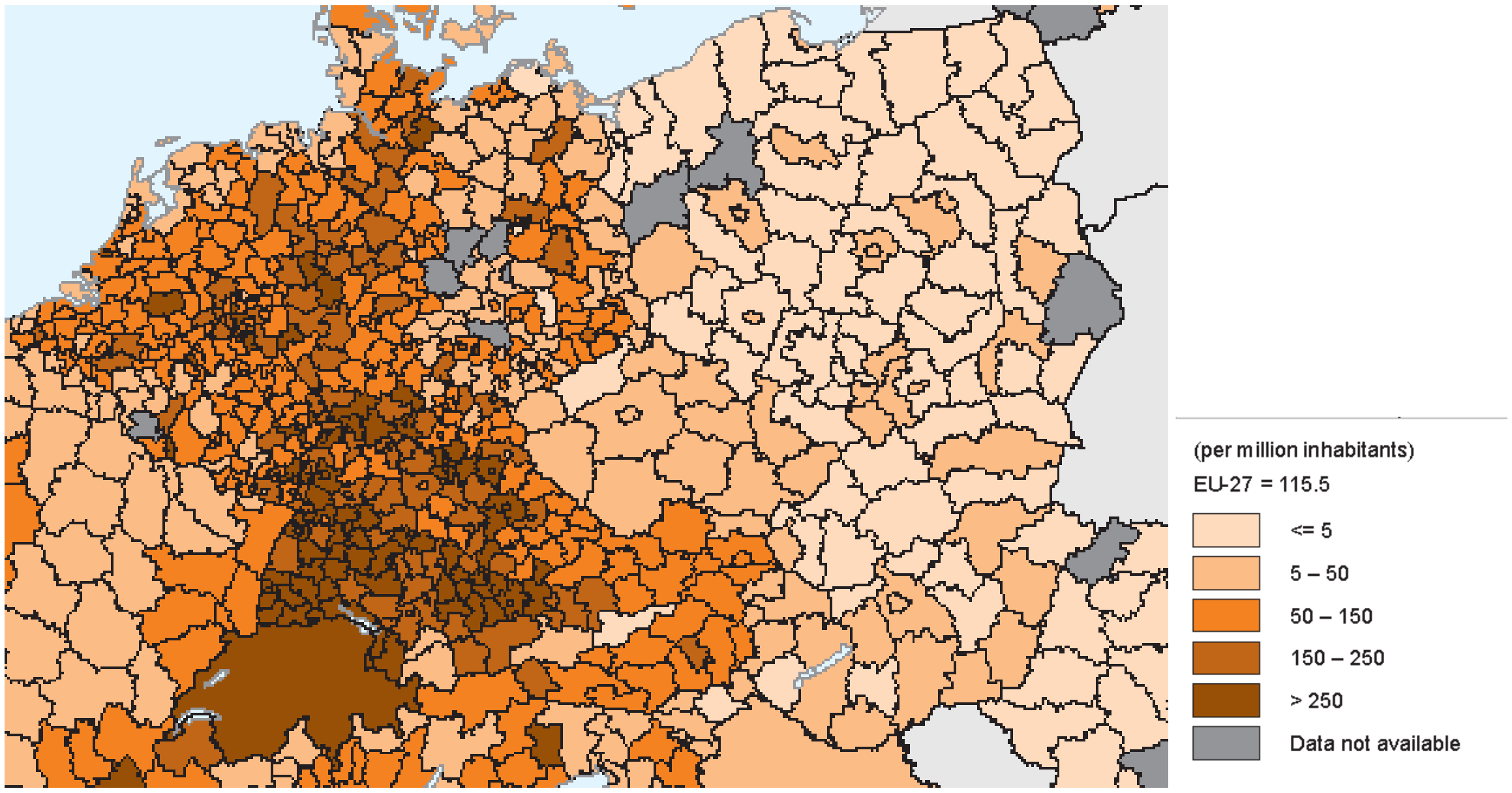

Source:
http://epp.eurostat.ec.europa.eu/statistics_explained/index.php?title=File:Patent_applications_to_the_EPO,_by_NUTS_3_regions,_2008_%281%29_%28per_million_inhabitants%29.png&filetimestamp=20120508162219

**Fig. 4:** Rank-size correlations of West Germany, East Germany, and CEE countries locations with respect to their patenting activity

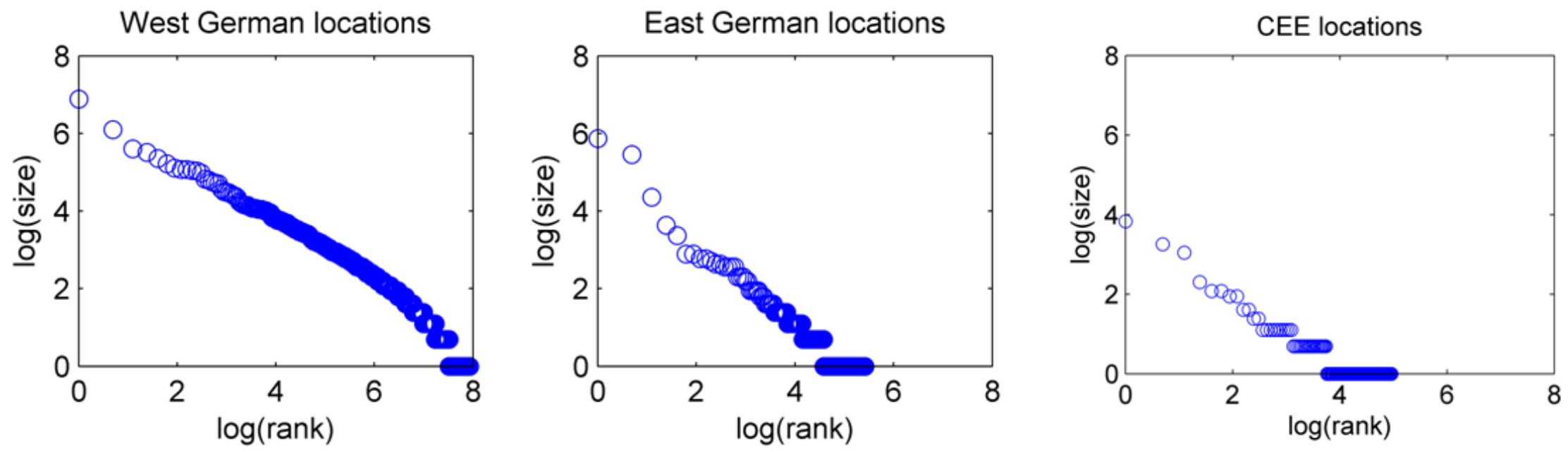

**Fig. 5** Rank-size correlations of West Germany, East Germany, and CEE countries locations with respect to their patenting activity controlled by population size

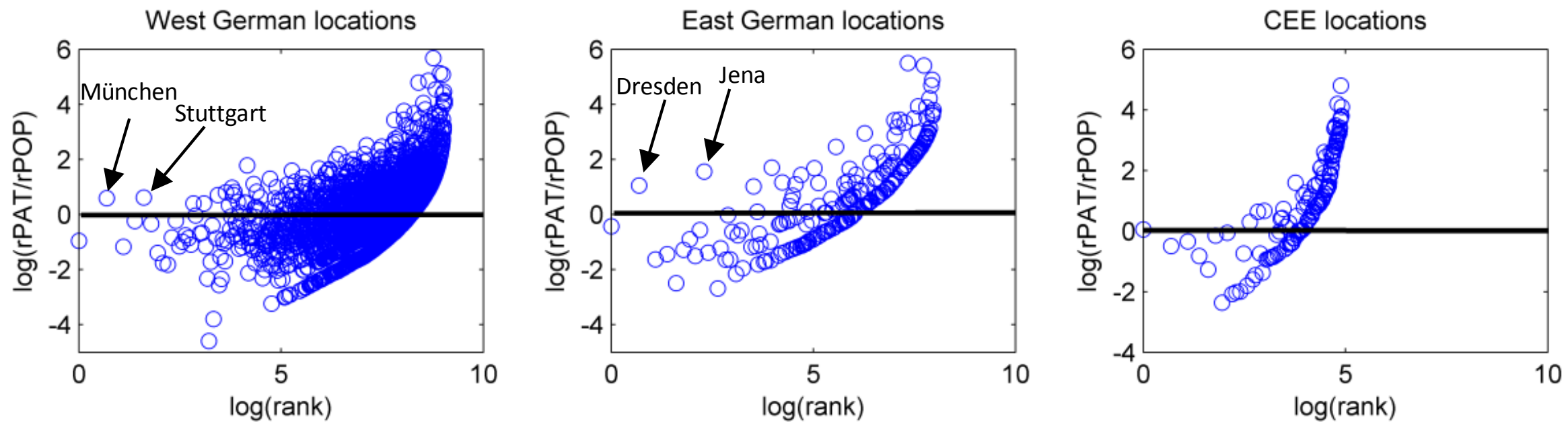

**Fig. 6:** Foreign ownership of USPTO patents in Germany, the Netherlands, and CEE countries

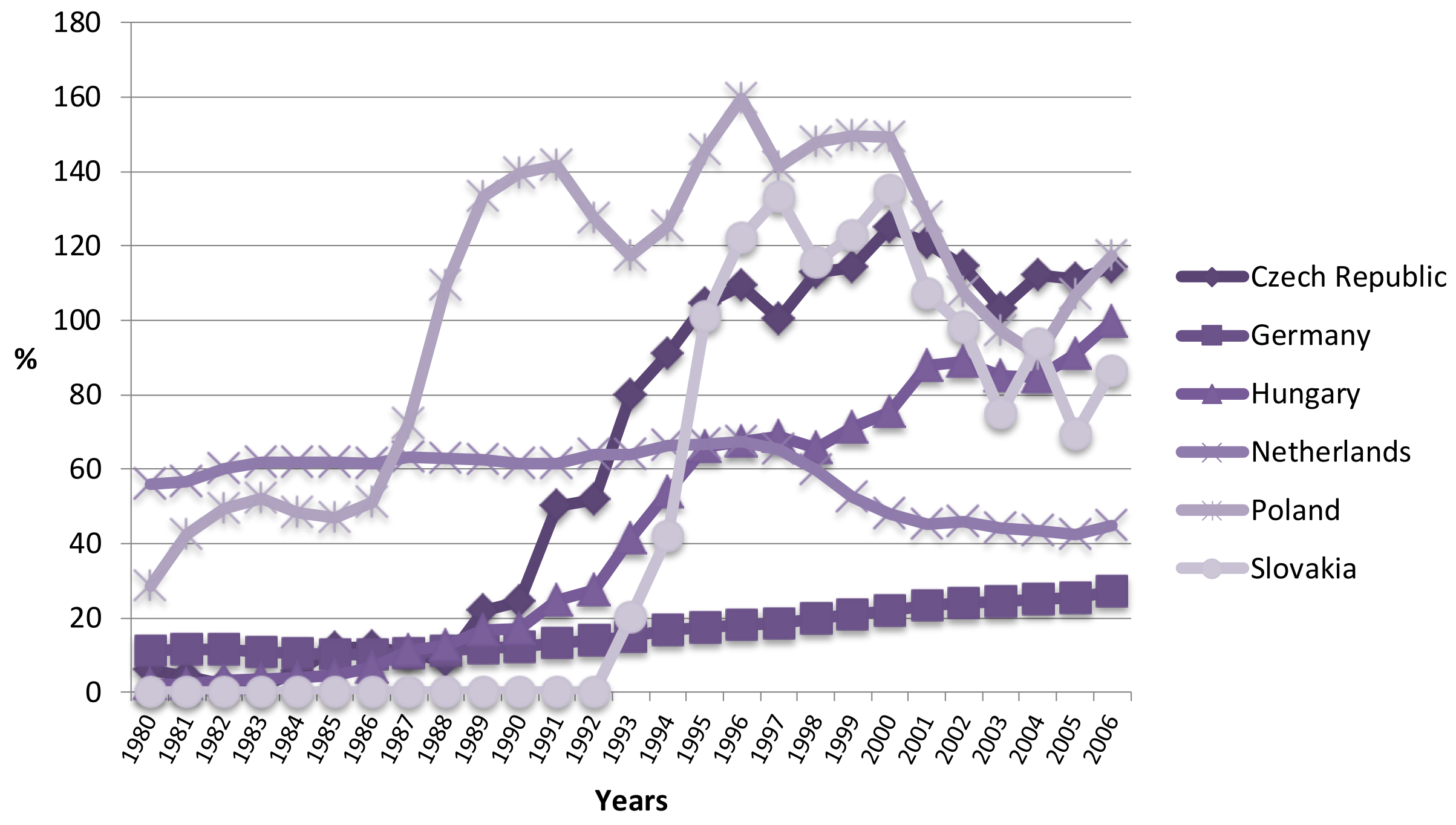

Source: self-edited from EUROSTAT data retrieved from
http://appsso.eurostat.ec.europa.eu/nui/show.do?dataset=pat_us_nfgn&lang=en
Note: Foreign ownership percentage represents the number of patents with at least one inventor from country X that has at least one assignee from country Y divided by the number of patents with at least one inventor from country X. Points represent three-year moving average of foreign ownership of values in the years *t-1*, *t*, *t+1*.

Appendix 1: Spatial evolution of USPTO patenting over the post-socialist transition, 1981-2010

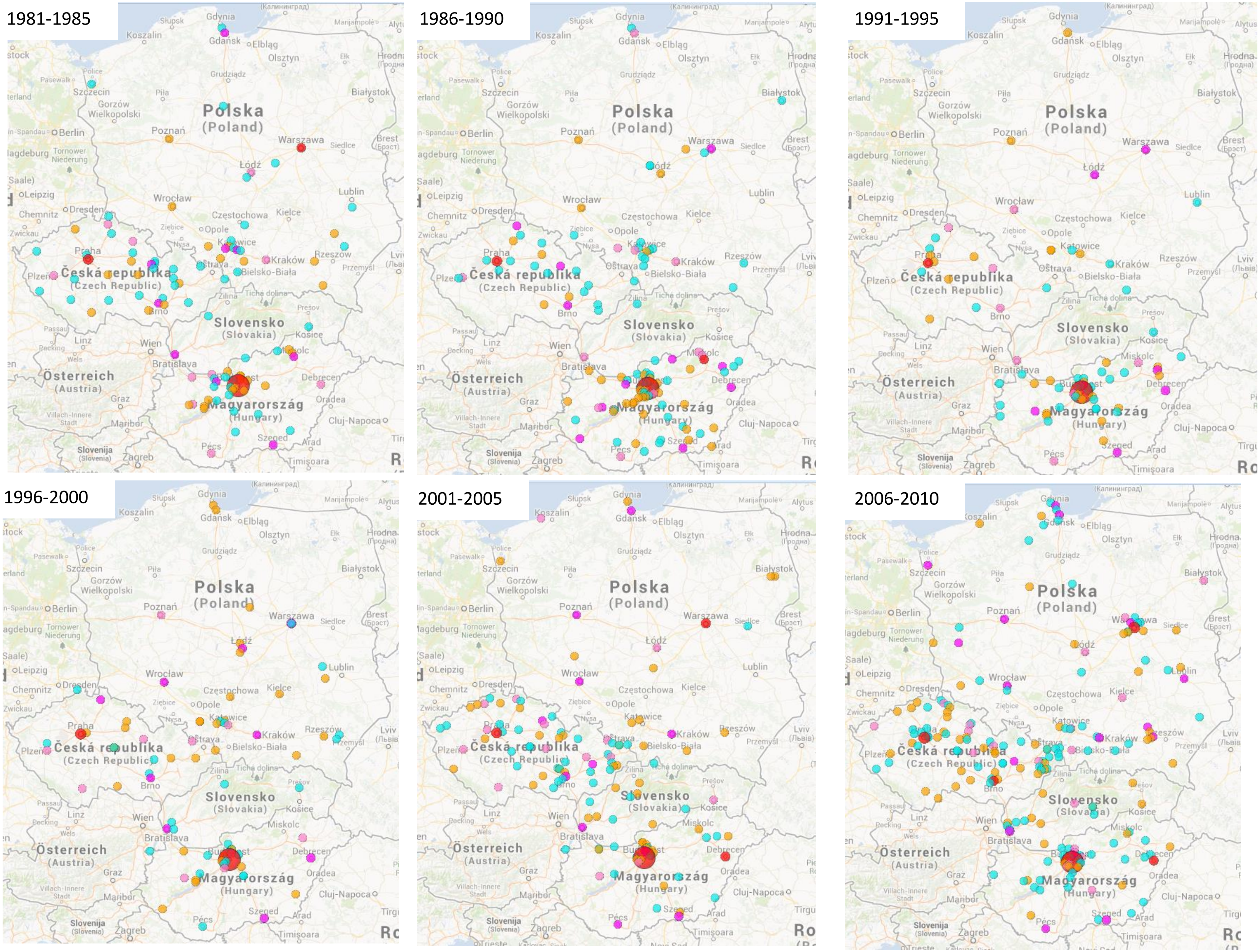

Appendix 2: List of companies from CEE countries with USPTO patents granted in 2007

| | Czech Republik | Hungary | Poland | Slovakia |
|---|---|---|---|---|
| SMEs | **IQI** (IT)<br>**Microrisc** (IT) | **Genoid** (biotechnology) | **Mectronic** (IT)<br>**Tokarz** (IT)<br>**Bury** (transport)<br>**Emporio spolka** (real estate) | **HighChem** (software engineering) |
| Big domestic companies | **Bran** (locking systems)<br>**Jihostroy** (hydraulics, aerospace engineering)<br>**Tescan** (electron microscope) | **Richter** (pharmaceuticals) | **Adamed** (pharmaceuticals)<br>**Seco/Warwick** (metal manufacturing) | **Dusto** (chemicals and agribusiness) |
| MNEs | **BSC Holice** (metal manufacturing)<br>**Uniplet** (machinery)<br>**Skoda** (vehicle)<br>**Zentiva** (pharmaceuticals)<br>**TRW DAS** (vehicle safety)<br>**Chemopetrol** (energy)<br>**RWE** (energy) | **EGIS** (pharmaceuticals)<br>**Teva** (pharmaceuticals)<br>**NABI** (vehicles)<br>**Tyco Electronics** (electronic machinery) | **ADB** (internet trade) | |